\documentclass{article}

\usepackage{graphicx}
\usepackage{booktabs}
\usepackage{latexsym}
\usepackage{amsmath}
\usepackage{hyperref}
\usepackage{pxfonts}
\usepackage{cancel}
\usepackage{bbding}
\usepackage{subfig}
\usepackage{subfloat}
\usepackage{caption}
\DeclareGraphicsExtensions{.pdf,.png,.jpg}

\title{Refraction of $e^-$ beams due to plasma lensing at a plasma-vacuum interface -- applied to beam deflection in a acceleration cell with electrical RF-breakdown plasma}

\author{A. A. Sahai \thanks{aakash.sahai@duke.edu}, T. C. Katsouleas \\ECE, Duke University, Durham, NC, 27708 USA}

\begin{document}

\maketitle

\begin{abstract}

We formulate a description of the deflection of a relativistic $e^-$ beam in an inhomogeneous copper plasma, encountered by the beam when propagating through a accelerating cell that has undergone a high electric-field RF-breakdown. It is well known that an inhomogeneous plasma forms and may last for up to a few micro-seconds, until recombination in an accelerating structure where a field-emission triggers melting and ionization of RF-cell wall deformity. We present a preliminary model for the beam deflection due to collective plasma response based upon the beam density, plasma density and interaction length.
 
\end{abstract}

\begin{figure}[ht!]
\center\includegraphics[angle=0,scale=0.5]{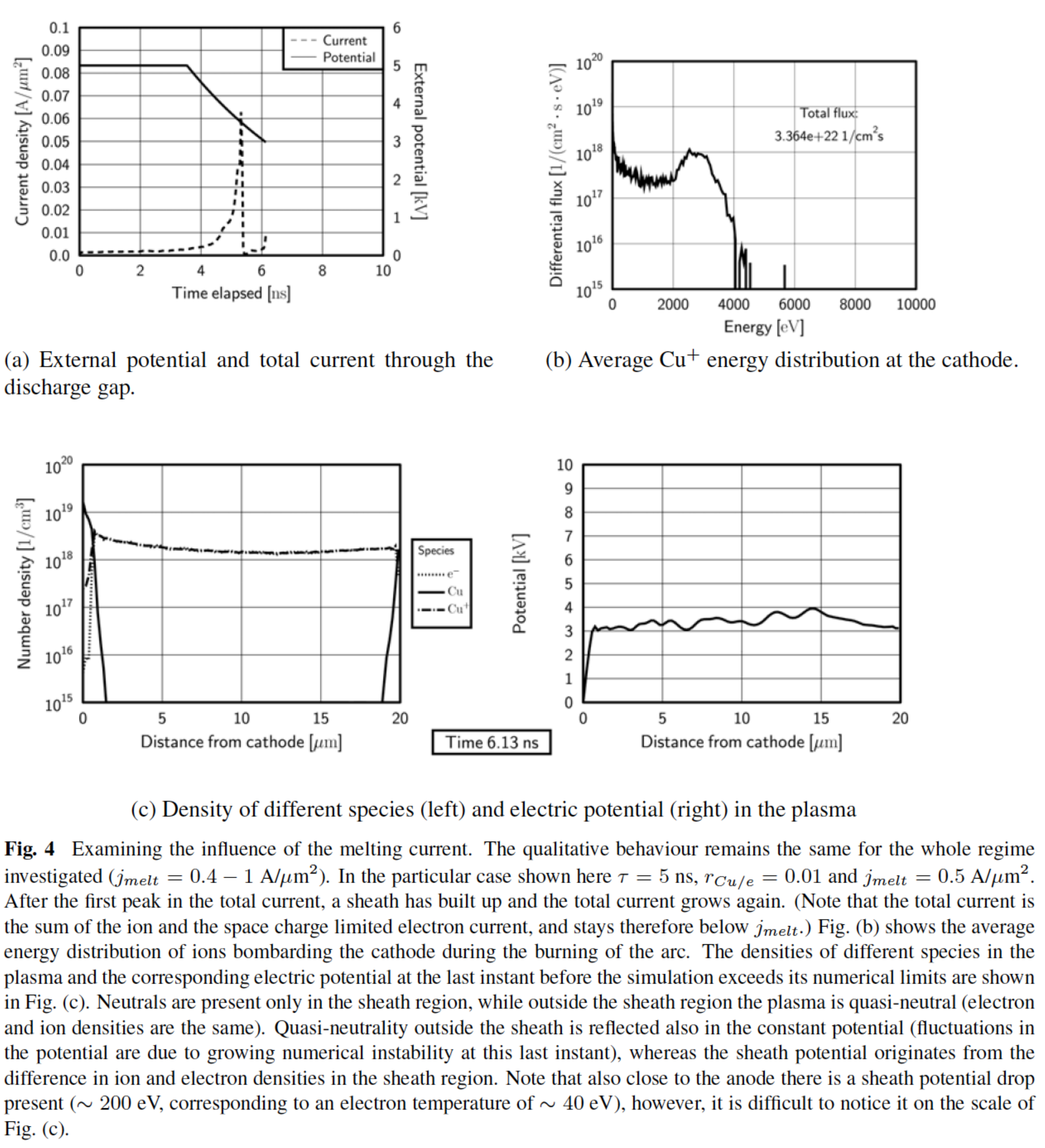}
\caption{Plasma density in a high electric-field breakdown cell simulated for a 20$\mu m$ gap \cite{timko-cpp-2011}}
\label{density-simulations-fig}
\end{figure}

We formulate a possible description of the deflection of a relativistic $e^-$ beam in copper plasma, encountered by the beam when propagating through a CLIC accelerating cell that has undergone a high electric-field breakdown\cite{CTF3-CERN}. Using the information from \cite{stahl-cern-2010}, we know that an inhomogeneous plasma forms \cite{timko-cpp-2011} (captured in Fig.\ref{density-simulations-fig}) and may last for up to a few micro-seconds, until recombination.\\

Just as a light beam bends {\it away from normal} when propagating from {\it a denser to rarer medium} (higher refractive index to lower index medium). Similarly, it has been proposed \cite{katsouleas-NIMA-2000} and experimentally observed \cite{muggli-Nature-2001} that a relativistic beam bends away from its propagating axis when exiting a plasma (denser plasma) into vacuum / neutral medium (lower plasma density).\\

As introduced in \cite{katsouleas-NIMA-2000}\cite{muggli-Nature-2001}\cite{muggli-PRSTAB-2001} the collective refraction of a relativistic beam in plasma is orders of magnitude larger compared to beam bending due to random scattering in a neutral medium. To demonstrate this the authors have used a comparison, a 30GeV beam through 1mm water cell bends due to random scattering by a rms-scattering angle of $20\mu rad$. The same beam going through a plasma / neutral (or vacuum) interface can undergo $1mrad$ of bending angle. Even though the plasma used is $10^{-7}$ times the density of water. This is due to the collective response of the plasma.\\

\begin{figure}[ht!]
\center\includegraphics[angle=0,scale=0.6]{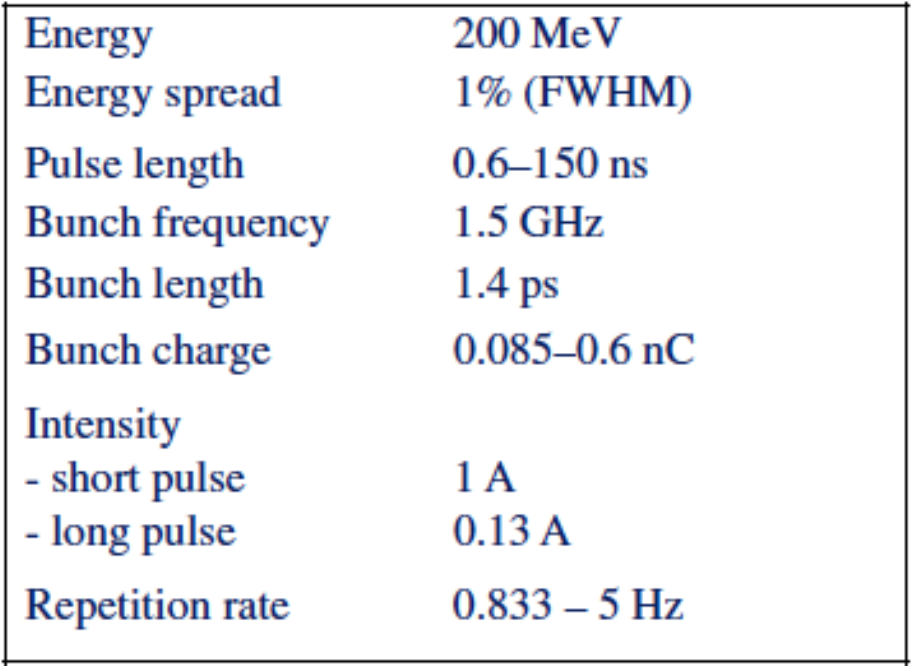}
\caption{CALIFES probe beam properties}
\label{CALIFES-beam-property}
\end{figure}

\section{Beam propagating in an inhomogeneous plamsa}
When an $e^-$ beam of charge volume density $n_b$ propagates through a plasma of density $n_0$ and $\frac{n_b}{n_0}\le1$ (over-dense), the beam $e^-$s space-charge fields at the head of the beam, repel the plasma-$e^-$s of the order of plasma electron density. The region depleted of plasma $e^-$s forms a region in the plasma with excess plasma ions. The displaced electrons allow the remaining ions to short out the space-charge electric field in the beam, but the magnetic pinch force of order of the beam space charge force remains and provides a net focusing force (or in the asymmetric case, a deflecting force). However, in an inhomogeneous plasma (plasma density changing spatially) modeled as plasma-vacuum interface (worst possible scenario), the density transition of the inhomogeneity is a step function. The beam experiences an {\it asymmetric plasma}, one half of the cylindrical beam is in the plasma whereas the other half is in the vacuum. Due to this imbalance of force on the beam due to the electron depleted region in the plasma, a deflecting force is applied on the $e^-$-beam, resulting in a net deflection.\\

The highest possible deflection angle, $\theta$ can be estimated form the impulse due to the focussing force of the electron depleted region, $F_{\perp}.\Delta t$. The electron depleted region (modeled as a half-cylinder of uniform charge equal to the beam charge) exerts a force on the beam, $F_{\perp}=-eE=2n_be^2r_b$ (Coulomb's law). The time spent within a CLIC cell composed of such an inhomogeneous plasma sustaining an electron depleted region due to beam electrostatic fields is $\Delta t=\frac{W}{c}$, where W is the width of a CLIC cell. The deflection angle can be estimated by $\frac{p_{\perp}}{p_{\parallel}}=\frac{F_{\perp}.\Delta t}{\gamma m_ec}$.

\begin{align}
\nonumber \theta&=\frac{2n_be^2r_b.\frac{\Delta W}{c}}{\gamma m_ec}\\
\nonumber &=2 \left(\frac{N}{\sqrt{2\pi} ~ \sigma_z \pi r_b^2}\right) r_b \frac{e^2}{\gamma m_ec} \left(\frac{\Delta W}{c}\right)\\
\nonumber &= 2 \frac{N}{\pi \sqrt{2\pi} ~ \sigma_z} \frac{e^2/m_ec^2}{r_b} \frac{\Delta W}{\gamma}\\
&\boxed{\theta_{inhomogeneous}= 2 \frac{N}{\pi \sqrt{2\pi} ~ \sigma_z} \frac{r_e}{r_b} \frac{\Delta W}{\gamma} }
\label{plasma-kick}
\end{align}

Taking the following parameters, modeled after data from fig.[\ref{density-simulations-fig}] and fig.[\ref{CALIFES-beam-property}] for CALIFES probe-beam.

\begin{align}
\nonumber N &= 0.2 \times 10^{10}\\
\nonumber \sigma_z &= 1.4 ~ ps \times c = 0.04 ~ cm\\
\nonumber \Delta W &= 0.15 ~ cm\\
\nonumber \gamma &= 392\\
\nonumber r_e &= 2.28 \times 10^{-13} ~ cm\\
& \boxed {r_b = 0.001 ~ cm} ~~ (assumed ~~ 10\mu m) 
\label{parameters-CLIC}
\end{align}

With these parameters we estimate a {\bf worst case deflection angle} $\theta_{inhomogeneous}$, for the asymmetric plasma channel, per CLIC cell of $\boxed {\simeq ~~ 137 ~ mrad}$.

\section{Beam propagating at an angle to the plasma-vacuum interface}

When an $e^-$ beam of charge volume density $n_b$ propagates through a (under-dense) plasma of density $n_0$ and $\frac{n_b}{n_0}>1$, the beam $e^-$s space-charge fields at the head of the beam, repel the plasma-$e^-$s out to $r_c=\alpha\sqrt{\frac{n_b}{n_0}}r_b$ \cite{whittum-prl-1990}, on each side of the axis of the beam (where $r_b$ is the beam radius). The region evacuated of $e^-$s forms an ion-channel, this exerts a focussing force upon the propagating beam. However, at the plasma-gas or plasma-vacuum interface, the ion-channel terminates leading to asymmetric fields on the beam, on either side of such an interface. Due to this {\it boundary effect}, a deflecting force is applied on the $e^-$-beam in addition to the focussing force, resulting in a net deflection at the boundary.\\

If the $e^-$ beam propagates at an angle of $\phi$ to the axis of the plasma. This deflection angle, $\theta$ can be estimated form the impulse at the boundary $F_{\perp}.\delta t$. The ion-channel (modeled as a cylinder of uniform charge) exerts a force on the beam, $F_{\perp}=-eE=2n_0e^2r_c$, which is calculated from Coulomb's law.  At the edges of the column there is an electron sheath with charge equal and opposite to the ion column, this sheath then repels the beam charge and causes the deflection. The time spent near the edge of the ion-column is $2\frac{r_c sin\phi}{c}$. The deflection angle can be estimated by $\frac{p_{\perp}}{p_{\parallel}}=\frac{F.\delta t}{\gamma m_ec}$. With these parameters we can estimate the deflection angle $\theta$.

\begin{align}
\theta_{interface} = \frac{1}{sin\phi}8\alpha\frac{N}{\pi\sqrt{2\pi}\sigma_z}\frac{r_e}{\gamma}
\label{angular-plasma-kick}
\end{align}

As it can be observed from \cite{muggli-Nature-2001}, the highest angle of deflection is when $\theta=\phi$. Also, the worst case scenario is when $\phi << 1 rad$, hence $sin\phi ~\simeq~ \phi$. So, the equation in eq.\ref{angular-plasma-kick} is modified as below.

\begin{align}
\boxed {\theta_{interface} = \sqrt{8\alpha\frac{N}{\pi\sqrt{2\pi}\sigma_z}\frac{r_e}{\gamma} }}
\label{angular-plasma-kick}
\end{align}

If we use the same parameters as in \ref{parameters-CLIC} for CALIFES probe-beam \cite{CTF3-CERN} with $\alpha=1$.

We estimate a deflection angle $\theta_{interface}$ of $\boxed {\simeq ~~ 6 ~ mrad}$.

\begin{figure}[ht!]
\center\includegraphics[angle=0,scale=1.0]{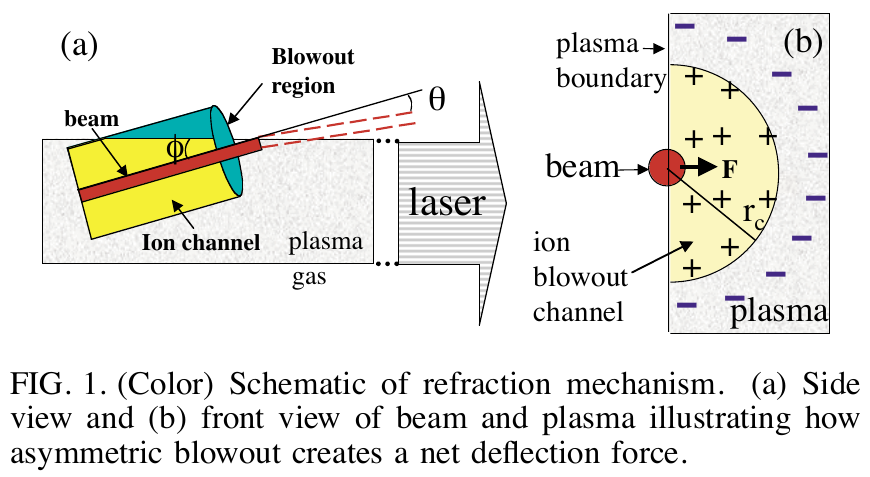}
\caption{Schematic of $e^-$-beam refraction at the exit of a plasma \cite{muggli-PRSTAB-2001}}
\label{beam-refraction-schematic}
\end{figure}

\end{document}